\pdfpageattr {/Group << /S /Transparency /I true /CS /DeviceRGB>>}
\documentclass[a4paper,twocolumn,noshowpacs,prl]{revtex4-1}
\usepackage{times,xspace}
\usepackage{amsbsy,amssymb,amsmath,bm}
\usepackage{graphicx,color,epsfig,rotate}
\usepackage{longtable}
\usepackage{rotating}
\usepackage{tikz}
\usepackage[T1]{sansmath}

\usetikzlibrary{arrows}

\def\bbbc{{\mathchoice {\setbox0=\hbox{$\displaystyle\rm C$}\hbox{\hbox
to0pt{\kern0.4\wd0\vrule height0.9\ht0\hss}\box0}}
{\setbox0=\hbox{$\textstyle\rm C$}\hbox{\hbox
to0pt{\kern0.4\wd0\vrule height0.9\ht0\hss}\box0}}
{\setbox0=\hbox{$\scriptstyle\rm C$}\hbox{\hbox
to0pt{\kern0.4\wd0\vrule height0.9\ht0\hss}\box0}}
{\setbox0=\hbox{$\scriptscriptstyle\rm C$}\hbox{\hbox
to0pt{\kern0.4\wd0\vrule height0.9\ht0\hss}\box0}}}}

\newcommand{\SiO}{{SiO$_2$}}
\newcommand{\BN}{{\emph{h}BN}}
\newcommand{\Vbg}{{$V_{\mathrm{bg}}$}}

\newcommand{\rxx}{{$\rho_{xx}$}}
\newcommand{\rxy}{{$\rho_{xy}$}}

\newcommand{\Bpara}{{$B_{\parallel}$}}
\newcommand{\Bperp}{{$B_{\perp}$}}

\begin{document}

\title{Rashba interaction and local magnetic moments in a graphene-Boron Nitride~heterostructure by
intercalation with Au}

\author{E.C.T. O'Farrell}
\email{ectof@icloud.com}
\affiliation{Centre for Advanced 2D Materials and Graphene Research Centre, National University of
Singapore, 117546, Singapore.}
\affiliation{Department of Physics, National University of Singapore, 117542, Singapore.}
\author{J.Y. Tan}
\affiliation{Centre for Advanced 2D Materials and Graphene Research Centre, National University of
Singapore, 117546, Singapore.}
\affiliation{Department of Physics, National University of Singapore, 117542, Singapore.}
\author{Y. Yeo}
\affiliation{Centre for Advanced 2D Materials and Graphene Research Centre, National University of
Singapore, 117546, Singapore.}
\affiliation{Department of Physics, National University of Singapore, 117542, Singapore.}
\author{G.K.W. Koon}
\affiliation{Centre for Advanced 2D Materials and Graphene Research Centre, National University of
Singapore, 117546, Singapore.}
\affiliation{Department of Physics, National University of Singapore, 117542, Singapore.}
\author{K. Watanabe}
\affiliation{National Institute for Materials Science, 1-1 Namiki, Tsukuba 305-0044, Japan.}
\author{T. Taniguchi}
\affiliation{National Institute for Materials Science, 1-1 Namiki, Tsukuba 305-0044, Japan.}
\author{B. \"Ozyilmaz}
\email{barbaros@nus.edu.sg}
\affiliation{Centre for Advanced 2D Materials and Graphene Research Centre, National University of
Singapore, 117546, Singapore.}
\affiliation{Department of Physics, National University of Singapore, 117542, Singapore.}


\begin{abstract}

We intercalate a van der Waals heterostructure of graphene and hexagonal Boron Nitride with
Au, by encapsulation, and show that Au at the interface is two dimensional. A charge transfer upon
current annealing indicates redistribution of Au and
induces splitting of the graphene bandstructure.
The effect of in plane magnetic field confirms that splitting is due to
spin-splitting
and that spin polarization is in the plane, characteristic of a Rashba interaction with magnitude
approximately 25 meV. Consistent with the presence of intrinsic interfacial electric field we show
that the splitting can
be enhanced by an applied displacement field in dual gated samples.
Giant negative
magnetoresistance, up to 75\%, and a field induced anomalous Hall effect at magnetic fields $<1$
T are observed. These demonstrate
that hybridized Au
has a magnetic moment and suggests the proximity to formation of a collective magnetic phase.
These effects persist close to room temperature.

\end{abstract}

\maketitle


Spin orbit coupling in graphene
is induced by hybridization with heavy metals \cite{weeks2011engineering}. This has been
achieved by intercalation of
graphene using e.g. Au, which produces a Rashba
interaction $\sim100$ meV \cite{marchenko2012giant}, and Pb \cite{calleja2015spatial}, on metallic
substrates. We
intercalate Au
into a heterostructure of graphene and dielectric hexagonal Boron Nitride (\BN), and report
spin-splitting of the graphene bands observed in quantum oscillations
on an insulating substrate, a crucial requirement for applications. The Rashba interaction
is large (25 meV) for samples intercalated with 0.1 monolayers (ML) of Au, but is modulated by
modest electric fields, thereby highlighting the requirement of hybridization
with spin-split Au
$d$-electrons \cite{marchenko2012giant}. We observe large negative
magnetoresistance, up to 75\%, indicating Au ions form magnetic
moments and an anomalous Hall effect suggests the possible formation of a collective magnetic phase.
The combination of Rashba interaction, magnetic moments and electric field
control of the density, is akin to dilute magnetic semiconductors \cite{dietl2014dilute}, and in a
Dirac material opens a route toward electric field control of magnetism and
engineering topological magnetic states such as the quantum anomalous Hall effect
\cite{qiao2010quantum,zhang2012electrically}.

\begin{figure}
	\begin{center}
		\includegraphics[viewport=185 445 425 775,clip=true]{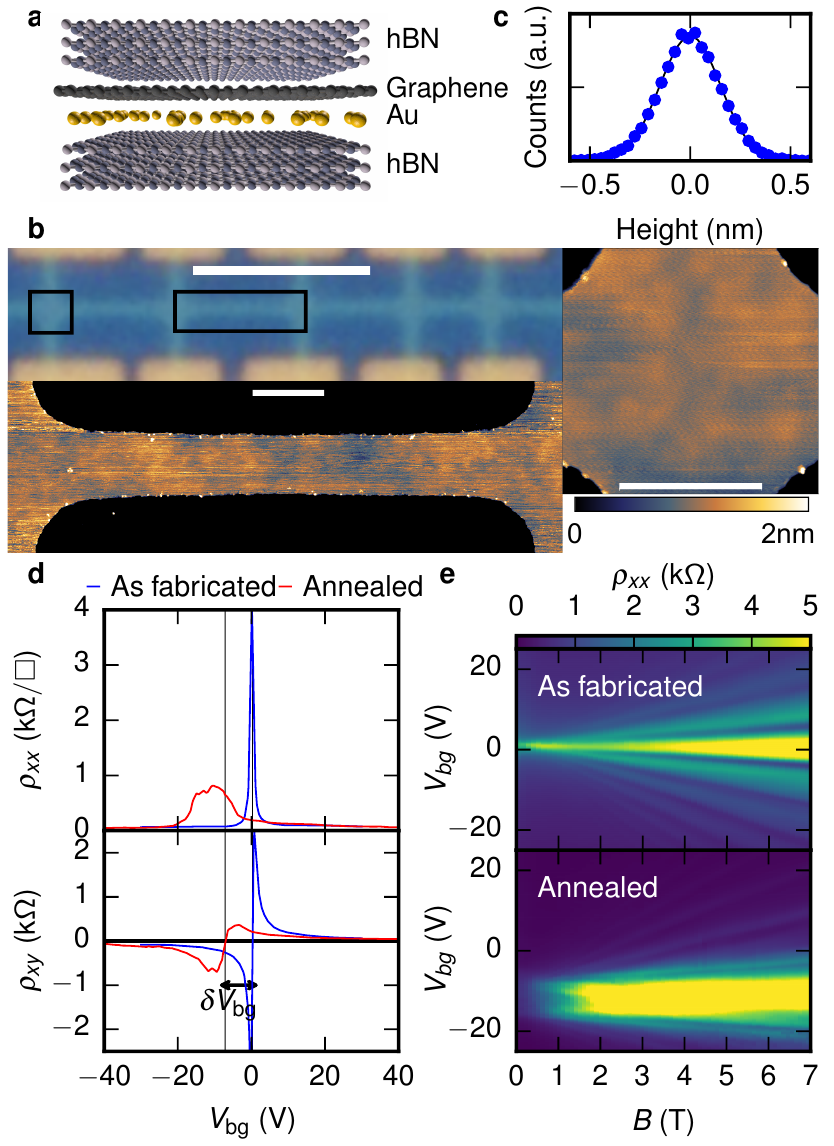}
	\end{center}
    \caption
    {\textbf{a}, Schematic of the Au intercalated
    graphene-\BN~heterostructure.
    \textbf{b}, optical (upper left, scale bar $5~\mu$m) and atomic force micrographs (scale bars
    $0.5~\mu$m) of a fully encapsulated and etched device, black boxes indicate regions in the
    atomic force micrographs.
    \textbf{c}, height histogram of a $1~\mu\mathrm{m}^2$ area of the device channel, fit to a
    Gaussian with roughness $=0.16$ nm.
    \textbf{d}, \rxx~ and \rxy~against \Vbg~at $T=1.2$ K and $B=0$ and 0.2 T, respectively, for a 0.1 ML
    intercalated device, both as fabricated and after current annealing.
    \textbf{e}, as fabricated and after current annealing \rxx~against \Vbg~and
    B, at $T=1.2$ K for a 0.1 ML intercalation.
}
	\label{fig:doping}
\end{figure}

The van der Waals interaction can create an atomically clean interface
between stacked two-dimensional (2D) crystals \cite{dean2010boron}; therefore, it can be considered
organizing principle for atoms or molecules
at a hetero-interface \cite{yuk2011graphene}. Here, we use the van der Waals interaction to cleanly
intercalate Au between graphene and \BN~and to induce strong
hybridization between graphene and Au. We intercalate graphene-\BN~heterostructures by depositing
0.1-0.5 nominal ML of Au
onto freshly cleaved \BN~on \SiO~in ultra high vacuum. Au decorated \BN~is used as the substrate onto which
a separately prepared graphene/\BN~(thickness $\approx20$ nm) is transferred, the structure is
illustrated in Fig.~\ref{fig:doping}a. The stack is etched and metallic
contacts are formed at the exposed edges of graphene \cite{wang2013one}.
Following thermal annealing the heterostructure is flat with root mean square roughness of
0.16 nm
(Fig.~\ref{fig:doping}b and c) over lateral regions of several micron, this
demonstrates the absence of 3D clusters of Au at the graphene/\BN~interface (see supplementary
information for additional information on cluster formation and Au motion). We fabricate the
device in these flat regions where
Raman spectroscopy measurements show the absence of bulk strain (supplementary information).


The interaction of Au and graphene has been extensively studied both by photoemission and transport.
Graphene on SiC intercalated with Au shows large shifts in doping ($>10^{13}$
cm$^{-2}$) depending on partial or full intercalation \cite{gierz2010electronic}, whereas graphene
intercalated on Ni remains
close to the neutrality point \cite{marchenko2012giant}. In transport measurements on
\SiO~quasi-continuous Au-film growth on
graphene induces weak $p$-doping of
$-5\times10^{11}~\mathrm{cm}^{-2}$, consistent with the lower vacuum
level of bulk Au. However, Au in a nanoparticle configuration \cite{wu2012tuning} induces $n$-doping
$+5\times10^{11}~\mathrm{cm}^{-2}$ attributed to the reduced Au-graphene separation and lower effective
potential.
Clustered Au produces local variations in the Fermi level and scattering leading to positive linear
MR \cite{jia2014large}
and an extrinsic spin Hall effect \cite{ferreira2014extrinsic,balakrishnan2014giant}.

Au is always present in the samples presented here, but as fabricated samples show longitudinal (\rxx)
and Hall (\rxy) resistivity
characteristic of pristine graphene, shown in Fig.~\ref{fig:doping}d for 0.1 ML Au and field effect
mobility
$1.2\times10^5~\mathrm{cm}^{2}/\mathrm{Vs}$; however, the MR
is negative (supplementary information). We observe Au at the interface is mobile even at room
temperature (supplementary
information), this suggests
isolated Au will gather into 2D clusters. To achieve the uniform Au distribution which is our goal
we perform a low temperature current anneal that redistributes and then quenches the Au configuration.
Low temperature post-growth annealing is typical in the synthesis of dilute magnetic semiconductors,
to evenly distribute magnetic ions
\cite{hayashi2001effect,li2007dopant}.

Current annealing (see supplementary information) induces diffusive
migration of Au on graphene
\cite{barreiro2011structured}. This
shifts the charge neutrality point by $\delta V_{\mathrm{bg}}=-8.4~\mathrm{V}$  making graphene
electron doped (Fig.~\ref{fig:doping}d), the doping is
$e\delta n_e=C_{\mathrm{bg}}\delta V_{\mathrm{CNP}}$, where $C_g$ is the
gate capacitance and $e$ is the electron charge, giving $\approx5\times10^{11}~\mathrm{cm}^{-2}$.
Following previous studies we
infer further $n$-doping is due to a less clustered, more uniform distribution. After annealing the
application of magnetic field $B$ shows that the Landau levels (LLs) are split
(Fig.~\ref{fig:doping}e). We note that the mobility of Au will lead to the loss of some Au
from the channel so the final concentration is less than deposition.


\begin{figure}
	\begin{center}
		\includegraphics[viewport=0 8 227 295,clip=true]{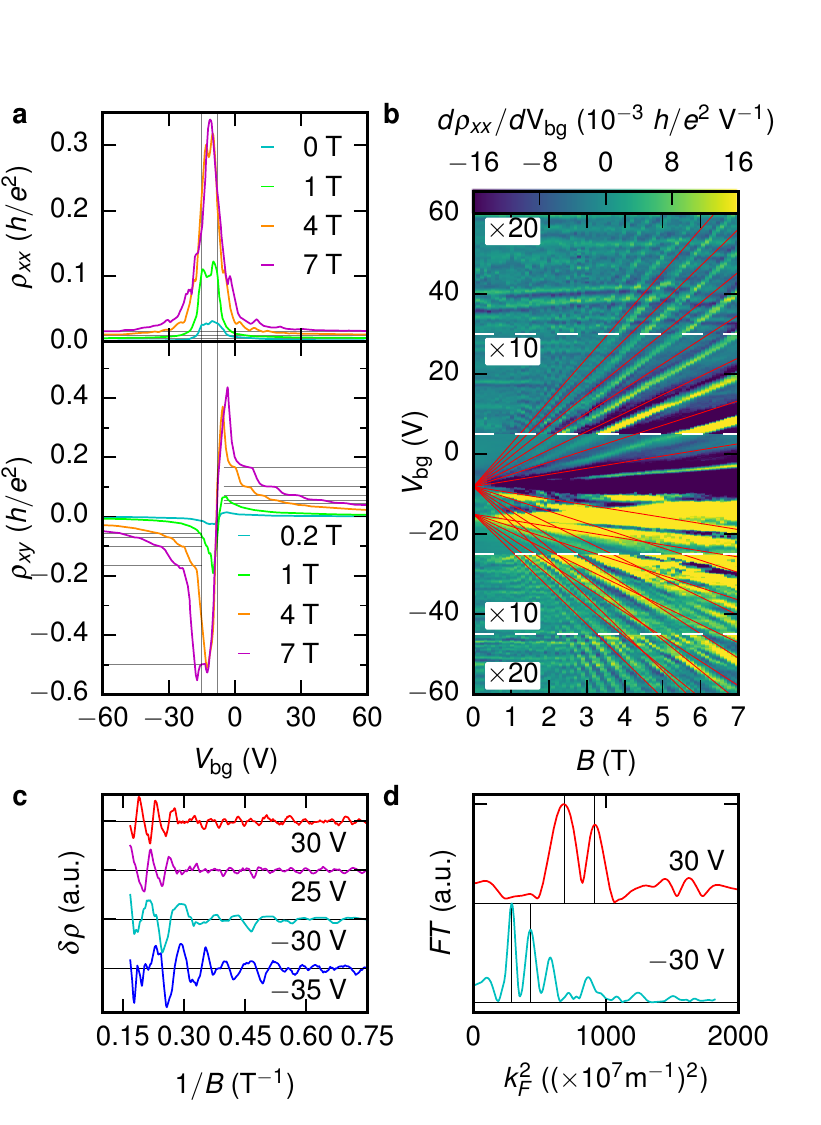}
	\end{center}
    \caption
    {\textbf{a}, \rxx~and \rxy~against \Vbg~for a 0.1 ML intercalated device at $T=1.2$ K, for
        various $B_\perp$ (\rxx~shifted by $0.01~h/e^2$ for clarity).
        \textbf{b}, derivative $d\rho_{xx}/V_{\mathrm{bg}}$~against
    \Vbg~and~$B_\perp$, overlaid with Landau fan as guide to the eye. Regions at positive and
    negative $V_\mathrm{bg}$ are magnified as indicated.
    \textbf{c}, oscillatory part of the MR against $1/B$ for various \Vbg.
    \textbf{d}, Fourier transform of the MR with first harmonic peaks indicated.
}
	\label{fig:2d}
\end{figure}

Figure~\ref{fig:2d}a shows the \rxx~against \Vbg~under various $B_\perp$. LL
formation is
visible, but, integer quantum Hall effect plateaus are visible only in \rxy. This is
comparable to the case of trilayer graphene which has both linear and quadratically dispersing bands
leading to a residual density of states
\cite{taychatanapat2011quantum}.

To resolve features in the electronic structure we take the
derivative $d\rho_{xx}/dV_{\mathrm{bg}}$ (Fig.~\ref{fig:2d}b). At positive \Vbg~well defined LLs are
clearly seen as part of a Landau fan from \Vbg$=-8.15$ V.
At positive \Vbg~we show a Landau fan for LLs with $N=\ldots,2,4,10,\ldots$ as a guide to the eye,
corresponding to the valley degeneracy of graphene.
At negative \Vbg~LL clearly emerge from a second Landau fan displaced by $\delta
V_\mathrm{bg}=-6.5$ V. This second fan has a pronounced asymmetry, appearing more prominently in the
valence band close to neutrality.
This asymmetry is sensitive to the concentration of Au intercalation, for
0.5 ML more pronounced splitting was observed in the conduction band (supplementary
information). Non-dispersive features are observed e.g. at \Vbg$\approx40$ V, these may arise from
damage to the contacts during annealing or extrinsic resonances
\cite{ferreira2014extrinsic}; however, these do not couple to the spin dependent
transport reported later.

The splitting wavevector is extracted from quantum
oscillations (Fig.~\ref{fig:2d}c and d) in the valence and conduction band. Clear nodes in
the oscillation are obscured by the presence of higher harmonic oscillations and anharmonicity.
Anharmonicity may arise from spatial variations in the splitting due to variation in the Au
distribution. We extract splittings
$k_R=3.8\pm0.2\times10^7$ and $4.0\pm0.2\times10^7$ m$^{-1}$ in the valence and conduction bands
respectively; yielding energy splittings $24\pm2$ and $26\pm2$ meV, respectively, at $E_F >100$
meV from neutrality, while splitting is stronger on the valence band side closer to neutrality. We now show
the splitting is spin splitting from a Rashba interaction.


\begin{figure}
	\begin{center}
        \includegraphics[viewport=190 395 425 692,clip=true]{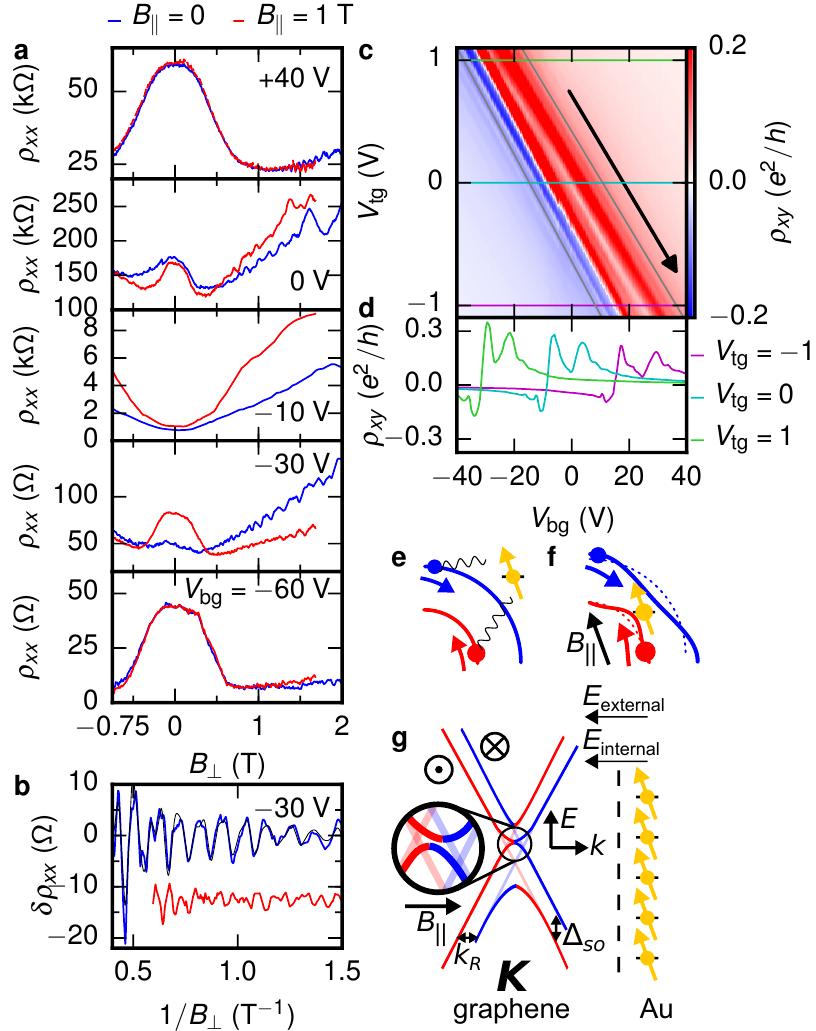}
	\end{center}
    \caption
    {
        \textbf{a}, \rxx~against \Bperp~ at several \Vbg, with and without \Bpara, at $T=1.2$ K.
        \textbf{b}, Oscillatory component of \rxx~for \Vbg$=-30$ V from \textbf{a}.
    \textbf{c}, effect of external electric field on \rxy~at $B=2$ T, $T=4.2$ K for a device with 0.5
    ML intercalation.
    Gray lines indicate contours of constant density in valence and conduction bands, arrow
    indicates direction of increasing external electric field.
    \textbf{d}, cross-sections through \textbf{c}.
    \textbf{e and f}, Schematic illustration of scattering processes of spin-split electrons.
    \textbf{g}, Schematic bandstructure of graphene with Rashba interaction, spin circulation
    anti-clockwise and clockwise indicated as red and blue. Direction of inferred internal electric
    field $E_{internal}$ and applied
    $E_{external}$ indicated. Expanded region indicates effect of \Bpara.
}
	\label{fig:InPlane}
\end{figure}

Fig.~\ref{fig:InPlane}a shows \rxx~versus \Bperp~at various \Vbg~(supplementary
information for additional \Vbg), with and without an
additional in plane magnetic field \Bpara$=1$ T.
Far from
neutrality there is negligible difference with and without \Bpara. Approaching neutrality
\Bpara~alters the MR; negative MR is enhanced and the LL structure is
significantly changed (Fig.~\ref{fig:InPlane}b). In the split region, $-15<$\Vbg$<-8$ V, negative MR
is absent and positive MR is enhanced.

The bandstructure of graphene with an Au induced Rashba
interaction \cite{rashba2009graphene,marchenko2012giant} is
illustrated schematically in Fig.~\ref{fig:InPlane}g for the $K$ valley, while spin polarizations are
reversed in the $K'$ valley. This identifies $-15<$\Vbg$<-8$ V as the regime where a single spin
band is occupied in each valley.
Negative MR is typical due to spin
flip scattering between spin polarized bands in materials with spin splitting, as
illustrated schematically in Fig.~\ref{fig:InPlane}e. Out of plane spin
fluctuations are suppressed under applied \Bperp and cannot mediate inter-band scattering so
resistance decreases. This is consistent with the results here
showing negative MR is absent in the single band regime.

In plane magnetic will affect these local moments, but the change in the LL spectrum
indicates instead changes in electronic structure.
In plane magnetic field is known to alter the electronic structure in spin-split bands due to the
effect of either finite quantum well width \cite{choi1988anisotropic,qu2015electric} or the Zeeman
interaction \cite{diez2015giant} as illustrated in Fig.~\ref{fig:InPlane}f. These typically occur in
relatively thick quantum wells
or large fields and this behavior is surprising in graphene which is
atomically thin and in pristine form has Land\'e $g$-factor $\approx2$
\cite{kurganova2011spin}, so the
Zeeman energy for \Bpara$=1$ T is $<1$ meV. We return to consider this point later.

Approaching the neutrality point from the valence band side negative MR
decreases when \Bpara$=0$ as the splitting wavevector increases. Negative MR
is enhanced by the application of \Bpara, suggesting that \Bpara~decreases the separation of spin
split bands in momentum space, illustrated in Fig.~\ref{fig:InPlane}f. In the
single band regime at -10 V the zero field resistance is not substantially changed but the MR is
significantly larger under \Bperp~for \Bpara=1 T. The strong positive MR in the single band regime
(Fig.~\ref{fig:doping}e)
may be the result of
perpendicular Zeeman interaction inducing a gap in the spectrum \cite{sodemann2015pc}, analogously
to the exchange
interaction \cite{qiao2010quantum}. This effect would be reduced by disorder but enhanced when the
bands are displaced
under \Bpara~as illustrated in the
expansion.

We now consider the effect of an additional applied electric field in a second sample with Au
concentration 0.5 ML and top and
bottom gates enabling the application of a displacement field to the Au/graphene interface. For
this sample splitting around the neutrality was found to be more strongly enhanced in the conduction band,
but graphene is again electron doped by Au. \rxy~at $B=2$ T is shown in
Fig.~\ref{fig:InPlane}d at several top gate voltages.

Au hybridized with graphene ionizes, donating charge to graphene and inducing an
electric field in the same direction. Even for the case of pristine graphene this
electric field can induce spin splitting \cite{min2006intrinsic,abdelouahed2010spin}; however, this requires
extremely large electric fields $>10$ V/nm.
Fig.~\ref{fig:InPlane}c shows that applying an external electric field the splitting can be enhanced
or reduced. When the external field is aligned with the internal field from Au to graphene (indicated by
the arrow) splitting increases such that contours of constant in the hole and electron bands are
driven apart,
indicated by the gray lines. Assuming energy and \Vbg~are approximately linear in this narrow
interval we observe 25\% increase in the splitting for an external field of $\approx0.5$
V/nm$^{-1}$. The relatively large effect of this modest external electric field therefore indicates that spin
orbit coupling is induced through spin dependent hybridization with the strong spin orbit coupled $d$-band of
Au deep in the valence band of graphene \cite{marchenko2012giant}.


\begin{figure}
	\begin{center}
		\includegraphics[width=\columnwidth,viewport=0 10 236 330,clip=true]{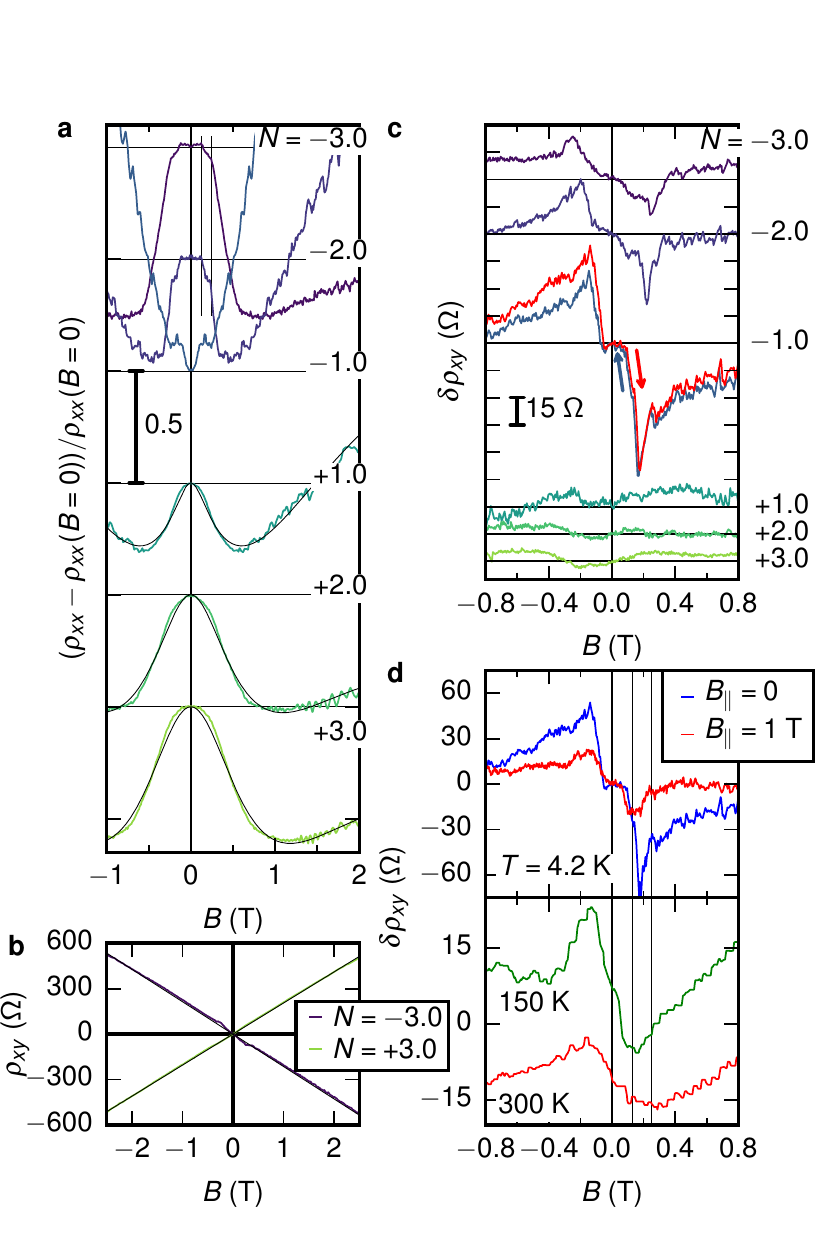}
	\end{center}
    \caption
    {Densities $N$ are $\times10^{12}$ cm$^{-2}$
        \textbf{a}, Low field MR at various $N$,
        $T=4.2$ K, dashed lines are fits to Eqn.~\ref{eqn:kf}.
    \textbf{b}, \rxy~at hole and electron densities, $T=4.2$ K.
    \textbf{c}, \rxy~after subtracting a linear normal Hall
    effect component.
    \textbf{d upper panel}, Effect of \Bpara~on non-linear component of Hall effect at
    $N=-1\times10^{12}~\mathrm{cm}^{-2}$. \textbf{Lower panel,} effect of temperature.}
	\label{fig:mr}
\end{figure}

We now consider the magnetic properties,
Fig.~\ref{fig:mr}a shows MR at various densities. At negative density sharp kinks are visible at
$B=0.11$ and 0.24 T, and the MR is up to 75\%. The absence of
hysteresis (supplementary information) at these fields indicates this is not a first order
transition such as ferromagnetism. At positive density the MR is
smooth and we fit the MR to the Khosla-Fischer expression for spin-flip scattering by local moments
\cite{khosla1970magnetoresistance}:
\begin{align}
    \begin{split}
    \frac{\Delta\rho}{\rho_0} & = -a^{2}\ln(1 + b^2 H^2) + \frac{c^2 H^2}{1 + d^2 H^2}\\
    a & = A_1JD(\epsilon_F)[S(S+1)+\langle M^2 \rangle]\\
    b^2 & = \Big[ 1 + 4S^2\pi^2 \Big(\frac{2JD(\epsilon_F)}{g}\Big)^4\Big]
    \Big(\frac{g\mu_B}{\alpha k_B T} \Big)^2 .
\end{split}
    \label{eqn:kf}
\end{align}
The first term is due to local moment scattering and the second is the MR of a 2
band metal; $a$ and $b$ are functions of the
magnetic exchange interaction $J$, the magnetization $M$, the local magnetic moment $S$, the
density at the Fermi level $D(\epsilon_F)$.
$A_1$ describes the contribution of spin scattering to the MR and $\alpha$ is a constant of assumed
to be of order 1. Eqn.~\ref{eqn:kf} fits well, the MR increases at higher density (see supplementary
information for
parameters). This implies Au acquires a magnetic moment.

Au is magnetic in various configurations, e.g. at interfaces
where magnetism is attributed to holes
in the $d$-band \cite{wilhelm2004magnetic}, but this mechanism
is unlikely to occur here where the Au $d$-levels lie several eV in the
valence band. $S$-electron magnetism
has been predicted in bare Au clusters \cite{luo2007s}, that may be particularly stable in 2D
\cite{choi2009neutral}. Experimentally magnetism is observed in bare Au nanoclusters and ascribed to
large spin-orbit coupling and reduced coordination at surfaces \cite{tuboltsev2013magnetism}.
We observe negative MR in the as fabricated samples, when Au is clustered, which is enhanced on
annealing suggesting the same mechanism may apply to moment formation here (supplementary
information). The moment size
cannot be independently extracted, but assuming it is
$\sim0.1~\mu_B$ implies that the graphene-Au exchange coupling is $O(\mathrm{eV})$, comparable to
dilute magnetic semiconductors \cite{dietl2014dilute}. 

Fig.~\ref{fig:mr}b shows \rxy~measured at hole and electron densities.
After subtracting a linear term for the normal Hall effect contribution
to obtain
$\delta\rho_{xy}=\rho_{xy}-R_H B$ we observe sharp kinks that are largest
at hole density, reaching approximately $90~\Omega$ at $N=-1\times10^{12}~\mathrm{cm}^{-2}$. These
features are reproduced between forward and reverse sweeps (red curve) without
hysteresis, we attribute these to the anomalous Hall effect (AHE).

The AHE arises from either intrinsic or extrinsic mechanisms. The extrinsic AHE
arises due asymmetric scattering of electrons from local moments in the presence of spin orbit
coupling. In Fig.~\ref{fig:mr}d we show that the application of \Bpara~decreases the AHE
contribution. The decrease in the AHE under
\Bpara~contrasts with the enhancement of negative MR by \Bpara
(Fig.~\ref{fig:InPlane}b); the decrease in $\delta\rho_{xy}$ when spin flip scattering is enhanced
indicates an intrinsic origin of the AHE. This AHE survives close to room temperature.

Further work is required to
characterize the magnetic behavior, but we speculate the AHE may occur due to chiral or topological
magnetic defects that tend to favored by the Dzyaloshinskii-Moriya interaction when inversion
symmetry is broken, such as here.
We note the similarity of the transport signatures observed here and those in topological magnetic
phases such as MnSi under
pressure \cite{ritz2013formation} and magnetic topological insulators
\cite{yasuda2016geometric}.

Finally we return to the large in-plane field effect on the electronic structure that is
most pronounced at low density, close to the single band regime. Spin splitting in graphene occurs
due to spin dependent hybridization with Au $d$-orbitals; this spin-orbit coupling also
drives the formation of local moments implying these phenomena are strongly coupled. Close to the
single band regime, where signatures of collective magnetism appear, the in-plane susceptibility of
the electron gas is strongly enhanced
\cite{zak2010spin}. This is likely to influence the magnetic state of the Au ions and thereby the spin splitting of
graphene. This effect may also be related to magnetoelectric effects proposed by Zhang et al.
\cite{zhang2012electrically}.

To summarize we have demonstrated electronic hybridization between graphene and Au in an
intercalated graphene/\BN~heterostructure. The observation of spin split LLs and the electric field
effect demonstrates a
Rashba interaction of approximately 25 meV, while negative MR and a large AHE suggest the
proximity to a collective magnetic phase. Our results
show that graphene hybridized with Au may be a platform for the electric field control of
interactions between magnetic moments, and could satisfy the proposed requirements to create
topological magnetic
phases \cite{zhang2012electrically}. Further work is required to fully elucidate the magnetic interactions
and enhance the size of these effects with other intercalation species.

\begin{acknowledgments}
We thank A.H. Castro-Neto, M. Cazalilla, A.R. Hamilton, T. Rappoport and I. Sodemann for fruitful
discussions and
S. Natarajan for assistance. This work was supported by the NRF, Prime Minister’s Office, Singapore,
under its Competitive Research Programme (CRP award number NRF-CRP9-2011-3), the SMF-NUS
Research Horizons Award 2009-Phase II and the NRF, Prime Minister Office, Singapore, under its
Medium-Sized Centre Programme.
\end{acknowledgments}

\end{document}